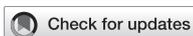

# Characterizing the hard and soft nanoparticle-protein corona with multilayer adsorption

Oriol Vilanova[1,2†], Alberto Martinez-Serra[3†], Marco P. Monopoli[3]* and Giancarlo Franzese[1,2]*

[1]Secció de Física Estadística i Interdisciplinària, Departament de Física de la Matèria Condensada, Universitat de Barcelona, Barcelona, Spain, [2]Institut de Nanociència i Nanotecnologia, Universitat de Barcelona, Barcelona, Spain, [3]Chemistry Department, Royal College of Surgeons in Ireland (RCSI), Dublin, Ireland

Nanoparticles (NPs) in contact with biological fluid adsorb biomolecules into a corona. This corona comprises proteins that strongly bind to the NP (hard corona) and loosely bound proteins (soft corona) that dynamically exchange with the surrounding solution. While the kinetics of hard corona formation is relatively well understood, thanks to experiments and robust simulation models, the experimental characterization and simulation of the soft corona present a more significant challenge. Here, we review the current state of the art in soft corona characterization and introduce a novel open-source computational model to simulate its dynamic behavior, for which we provide the documentation. We focus on the case of transferrin (Tf) interacting with polystyrene NPs as an illustrative example, demonstrating how this model captures the complexities of the soft corona and offers deeper insights into its structure and behavior. We show that the soft corona is dominated by a glassy evolution that we relate to crowding effects. This work advances our understanding of the soft corona, bridging experimental limitations with improved simulation techniques.

## 1 Introduction

Nanotechnology is a rapidly growing industry with emerging applications across various fields. Although nanostructures have been present in human life for a long time (Freestone et al., 2007), the understanding and development of advanced nanomaterials are relatively new (Qiu et al., 2017). In particular, there is an increasing interest in comprehending the behavior of nanoparticles (NPs) within biological systems (Dawson and Yan, 2021). The unique size, structure, and chemical properties of NPs introduce a wide range of new applications in many areas of research and technology, including therapeutics and diagnostics (Trinh et al., 2022).

When NPs are introduced into a biological environment, they quickly become coated by surrounding biomolecules, such as proteins, unless specifically designed not to do so (Cedervall et al., 2007b; Cedervall et al., 2007a; Lynch and Dawson, 2008). Research has shown that it is not the NP itself, but rather the biomolecules on its surface, that determine its interaction with living organisms. The layer of adsorbed proteins is known as the *protein*

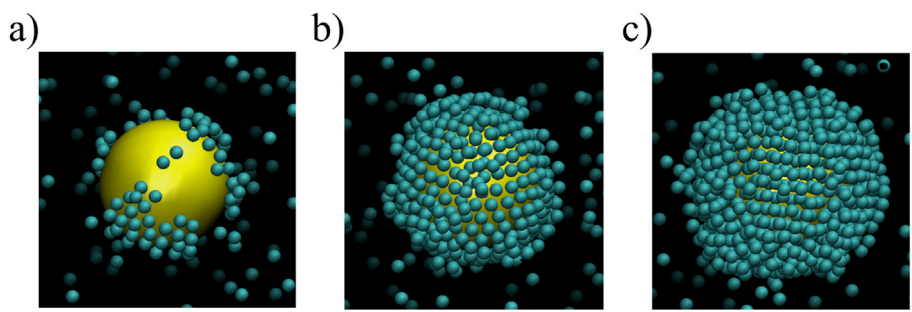

FIGURE 1
Coarse-grained configurations of Tf in suspension with polystyrene NPs. The relative concentration is [Tf]/[NP] = 2000 and [NP] = 1 mg/mL. The panels are snapshots taken at simulation times increasing from left to right. **(A)** Early-time formation of the hard corona. **(B)** Intermediate-time partial formation of the hard and the soft corona. **(C)** Large-time full formation of the hard and the soft corona.

*corona* and defines the biological identity of the NPs (Lynch et al., 2009; Walczyk et al., 2010; Monopoli et al., 2012).

Due to their size, NPs can distribute throughout organisms, reaching various cellular and organic compartments, and, in specific cases, they can even breach biological barriers, such as the blood-brain barrier (Kreuter et al., 2002; Wagner et al., 2012). Indeed, the size and shape of nanoparticles can influence the formation of the protein corona, and their effects on the overall protein composition remain a subject of debate (Lundqvist et al., 2008; Tenzer et al., 2013; Talamini et al., 2017; Xu et al., 2018; Madathiparambil Visalakshan et al., 2020; Bilardo et al., 2022). In medicine, NPs can serve as therapeutic tools, enhancing previous techniques like drug delivery as they extend overall circulation time and the drug's efficacy (Kumari et al., 2016), with the corona influencing the therapeutic outcome (Salvati, 2024). However, concerns have also been raised about the safety of prolonged or chronic exposure to NPs and the potential role of the corona in triggering the immune response (Oberdörster et al., 2005; Deng et al., 2011; Savolainen et al., 2013). Therefore, given the potential of nanotechnology, it is crucial to understand whether it poses a threat to organisms and the environment to ensure safe clinical translation of these new biomedical nanotools (Corbo et al., 2016; Boselli et al., 2024).

Numerous experiments have been conducted to understand how blood plasma proteins bind to a nanoparticle. The findings indicate that proteins adhere to the surface of the NP, forming a corona around it, consisting of two different components (Milani et al., 2012): a *hard* corona (HC) comprising tightly bound proteins in direct contact with the surface, and a more dynamic *soft* corona (SC) that is in constant exchange with the protein solution (Figure 1).

Experiments show that the HC binds almost irreversibly to the NP, while the SC binds reversibly (Lynch et al., 2009; Milani et al., 2012). The HC is considered the most biologically relevant part as it interacts with cells and biological machinery via the receptors. On the other hand, the SC plays a role in mediating transient and dynamic interactions with biological systems (García-Álvarez and Vallet-Regí, 2021; Bai et al., 2021). Characterizing the SC is challenging due to its transient nature and weak binding affinities, but recent advances help to understand its role. For example, using cryo-transmission electron microscopy (cryo-TEM) combined with synchrotron radiation circular dichroism (CD) allows real-time *in situ* insights into the transient nature and dynamic evolution of the SC. This approach reveals how weakly bound proteins within the SC undergo continuous exchange and reorganization (Sanchez-Guzman et al., 2020).

Also, *in situ* click-chemistry is used to map and identify the proteins within the SC, enabling precise tracking of protein-NP interactions. This technique reveals that the dynamic and reversible nature of the SC proteins significantly influences NP uptake and interaction with cellular membranes (Mohammad-Beigi et al., 2020).

Furthermore, the analysis of the protein corona on gold NPs by Liquid Chromatography-Tandem Mass Spectrometry (LC-MS/MS) and Enzyme-Linked Immunosorbent Assay (ELISA) has found that only 27% of the adsorbed proteins were functional for binding (Zhang et al., 2020). This indicates that the corona, rather than a simple monolayer, is an assembly of layers comprising a foundation layer (i.e., the HC) and an intermediate corona (IC) plus a binding layer, with the last two usually identified as SC.

Recent studies have combined different experimental methods to characterize $MoS_2$ nanomaterials *in vivo* and have demonstrated that the SC mediates in the biodistribution, transformation, and bioavailability of nanoparticles, thus impacting liver metabolism and enzyme activity (Cao et al., 2021).

Additionally, a "fishing" method has been developed using bio-layer interferometry and LC-MS/MS to monitor the dynamic formation and evolution of the protein corona on chiral $Cu_2S$ nanoparticles (Baimanov et al., 2022). This technique allows for real-time tracking and detailed characterization of the SC.

To understand how different layers form on the surface of NPs, particularly the SC, it is essential to characterize the kinetics of protein adsorption into the corona. The rate at which proteins are adsorbed and the protein corona formation depend greatly on protein concentrations and affinities. Competition and cooperation between different types of proteins are crucial mechanisms in understanding kinetic processes, such as the Vroman effect (Vroman and Adams, 1969; Vroman and Adams, 1986; LeDuc et al., 1995; Vilaseca et al., 2013). Only biomolecules that reside in the protein corona for longer than the characteristic timescale of a biological phenomenon are likely to be relevant to the process. Experiments have shown how tissues, organs, and other biological systems respond depending on how long biomolecules stay at the NP surface (Tran et al., 2017). Therefore, it is reasonable to assume that proteins with a long residence time near the surface are the ones that give the NP its biological identity. On the other hand, the proteins that attach to the NP only temporarily, depending on the current environment, may not be relevant.

These two different timescales suggest a memory effect on the NP. Exposing the NP to a new environment with different biomolecules could result in a partially new corona composition. Biomolecules that have not been replaced in the protein corona serve as a record of previous environments visited by the NP. This was studied in an earlier work, where it was demonstrated, with experiments and simulations, the memory effect on silica NPs suspended in solution with Human Serum Albumin (HSA) and Transferrin (Tf) using simulations and experiments (Vilanova et al., 2016).

Numerical simulations can provide valuable insights into the multilayer adsorption of proteins in the corona. However, they represent a challenge when compared with the experiments. Indeed, experimental techniques like Fluorescence Correlation Spectroscopy (FCS), used to quantify protein adsorption, typically involve very low concentrations of NPs (Rusu et al., 2004; Milani et al., 2012). At the same time, in biological cases of interest, protein concentrations are often very high, as in the case of SDS-PAGE (Sodium Dodecyl Sulfate-Polyacrylamide Gel Electrophoresis) experiments (Lundqvist et al., 2011; Pitek et al., 2012).

These conditions correspond to the worst-case scenario in molecular simulation. Indeed, low NP concentrations require extensive systems, eventually made of a large simulation box containing only a single NP. However, the need for high protein concentrations leads to an exponential increase in the number of proteins, making it demanding to simulate the processes. Furthermore, the problem becomes more challenging when simulation time scales need to match actual experiments' time frames, typically in minutes or even hours.

To achieve significant results within a specific timescale, it is crucial to select an appropriate simulation scheme, which is determined by the integration time step and the level of coarse-graining used in the simulation technique. Indeed, a full-atom description of thousands of interacting proteins is out of the present reach of even the most powerful computer clusters. Therefore, researchers resort to coarse-grained approaches that reduce the degrees of freedom involved in the biological systems at the nanoscale and simplify the description of the relevant interparticle interactions. Meaningful information can still be obtained using a minimal system description focusing on the most appropriate degrees of freedom.

While a general description of the solvent is essential for many biological processes (March et al., 2021; Durà-Faulí et al., 2023), the effects of hydration on protein interactions can be accounted for using effective potentials, especially when the model's transferability at different thermodynamic conditions is not crucial. Additionally, employing the Langevin integration scheme for the equations of motion allows for the simulation of the thermal energy contribution from water (Vilanova et al., 2016). In this simplified description, proteins are the primary focus, and their number influences the simulation dimensions.

Better computational performances can optimize numerical simulations, and parallelization is a straightforward way to achieve this goal. Over the past decade, Graphical Processing Units (GPUs) have emerged as a cost-effective and computationally efficient choice for implementing molecular simulations (Walters et al., 2008; Anderson et al., 2008; Harvey et al., 2009; Trott et al., 2010). Hence, the codes we describe and use in the following are developed within the CUDA® framework, which is compatible with most NVIDIA® GPUs.

Specifically, we developed BUBBLES ("BUBBLES is a User-friendly Bundle for Bio-nano Large-scale Efficient Simulations"), a suite of tools designed for simulating the interactions and kinetics of NPs in biological environments, modeled as aqueous solutions containing proteins (Vilanova, 2015). Here, as a case study, we use BUBBLE to analyze the SC kinetics of Tf adsorbing onto polystyrene (PSCOOH) NPs.

Polystyrene is a common polymer used in the production of plastic, such as in packaging materials, food containers, and disposable cups. These NPs have a low density and high surface area-to-volume ratio, making them useful in applications like electronics or biomedical research, despite health concerns arising (Kik et al., 2020).

On the other hand, Tf is one of the most abundant proteins in human plasma (Schenk et al., 2008). It is a glycoprotein with a molecular weight of around 80 kDa, composed of two subunits of equal size joined by a disulfide bridge. Each subunit has a single site for binding iron (Gomme et al., 2005). Iron atoms are absorbed in the intestine, bound to Tf in the plasma, and then transported to storage and utilization sites such as the bone marrow and the liver (Gkouvatsos et al., 2012).

We aim to describe the results observed in experiments involving polystyrene NPs in solutions containing Tf (Milani et al., 2012) and to predict by numerical simulations of our coarse-grained model the short-time dynamics of the SC formation. Next, to validate the model's applicability in protein-rich environments, we use Differential Centrifugal Sedimentation (DCS) to assess NPs size distribution after exposure to Tf.

## 2 Materials and methods

When simulating biological systems interacting with nanoscale objects using atomistic simulations, handling many proteins for long simulation time scales is prohibitive. Additionally, a significant amount of resources are needed to simulate the solvent. To address this, we developed a coarse-grain approach that provides a simplified system description.

This method retains all the essential molecular details, making it suitable for studying bio-nano interactions at the mesoscale. Our approach is based on an implicit solvent description, which uses phenomenological parameters and significantly reduces the computational cost.

However, the strong approximations in the model make it not transferable to different thermodynamic conditions or protein-NP combinations. To adjust the model's parameters, preliminary experiments are necessary to measure the adsorption isotherms of each protein onto the NPs of interest. Nevertheless, these experiments are easily manageable and represent no impediment to calibrating the model's parameters at the desired thermodynamic condition.

## 2.1 Experimental details

### 2.1.1 Materials

Polybead® Carboxylate Microspheres 0.10 μm (PSCOOH, nominally 100 nm) were purchased from Polysciences Inc. (Warrington, United States). The colloidal stability of the nanoparticles was ensured by measuring their size distribution in PBS before protein exposure. Phosphate-Buffered Saline (PBS) tablets, D-(+)-sucrose (99.9%) and Human Transferrin (T8158) were purchased from Sigma Aldrich Ireland. We dissolved one PBS tablet in 200 mL of ultrapure water to obtain a 0.01 M phosphate buffer, 0.0027 M potassium chloride and 0.137 M sodium chloride solution with a pH of 7.4 at 25°C.

### 2.1.2 Preparation of the NP-Corona complexes

The protein corona samples were prepared using protocols previously developed in the lab (Soliman et al., 2024b). Different volumes of PBS were added to low protein binding 1.5 mL microtubes to make the final solution volume 1 mL. Then, various volumes from a 10 mg/mL Tf stock were added to the microtubes to achieve different experimental concentrations (0, 400, 675, 950, and 1,500 [Tf]/[NP]). After that, 7.5 μL from a stock solution of polystyrene microspheres was added to reach a final NP concentration of 1 mg/mL. The solution was then directly injected into the analytical centrifuge.

### 2.1.3 NP physico-chemical characterisation

Differential Centrifugal Sedimentation (DCS) experiments were conducted using a CPS Disc Centrifuge DC24000 (Analytik Ltd.) with a sucrose gradient ranging from 8% to 24%. We used polystyrene NPs with a diameter of 0.522 μm to calibrate each sample measurement. The travel time of spherical particles with uniform density from the disk's center to the detector correlates directly with their size. Variations in arrival times allow for the differentiation between populations, which are considered *apparent* sizes. We calculated the shell thickness using a core-shell model with a protein layer density estimated to be 1.15 g/cm³ (Perez-Potti et al., 2021). Dynamic Light Scattering (DLS) and Zeta Potential measurements were performed using Zetasizer Nano ZS (Malvern). The sample cuvettes were equilibrated at 25°C for 90 s. For each measurement, the number of runs and duration were automatically determined and repeated three times. Measurements and data analysis were performed according to standard procedures (Soliman et al., 2024b).

## 2.2 Model

For the case of interest here, we simulate the materials and proteins used by Milani et al. (2012). Specifically, we coarse-grain the NPs as spheres with radius $R_{NP}$ = 35 nm, corresponding to the hydrodynamic radius estimated by dynamic light scattering (DLS) of the NPs used by Milani et al. in Milani et al. (2012), the polystyrene (PSCOOH) density $\rho_{PS}$ = 1.05 g/cm³, and mass $M_{NP}$ = 1.08 × 10⁵ kDa, in a solution containing Tf proteins[1]. In a single-protein solution such as the one considered here, there is no competitive binding among different types of proteins surrounding the NP, allowing us to gain an accessible insight into the dynamics of the SC formation.

We represent the Tf as globular proteins described by coarse-grain parameters as in Vilanova et al. (2016). Specifically, Tf has a mass $M_{Tf}$ = 80 kDa and a hydrodynamics radius coinciding with the one calculated based on its maximum surface concentration upon adsorption, $R_{Tf}$ = 3.72 nm (Milani et al., 2012; Pitek et al., 2012). Although here we focus on Tf, a protein that closely resembles a spherical shape, the model also effectively addresses the distinctly non-spherical structure of, e.g., fibrinogen, as discussed in previous works (Vilaseca et al., 2013; Vilanova et al., 2016). Furthermore, our approach also allows the model to account for the different conformational states of proteins during adsorption and how they impact binding kinetics (Vilaseca et al., 2013; Vilanova et al., 2016).

### 2.2.1 Particle interactions

We adopt an implicit water approach to account for the water's effects, with effective interaction potentials between the proteins and the proteins and the NP. Our experiments show that Tf does not aggregate under our experimental conditions. Therefore, we model the effective interaction between two proteins in solution as a soft-sphere repulsive potential, as described in Equation 1

$$U_{PP}(r_{ij}) \equiv \left[\frac{2R_{Tf}}{r_{ij}}\right]^{24} \quad (1)$$

where $R_{Tf}$ is the Tf radius and $r_{ij}$ is the center-to-center distance between the proteins $i$ and $j$.

As demonstrated in Vilanova et al. (2016), the protein-NP interaction can be described within the framework of the Derjaguin, Landau, Verwey, and Overbeek (DLVO) theory for charged solutes in a solvent, whose stability is controlled by the balance between van der Waals attraction and electrostatic interaction, often leading to a short range electric double-layer repulsion, and a strong contact attraction (Derjaguin and Landau, 1993; Verwey et al., 1948; Agmo Hernández, 2023). In the context of adsorption problems, instead, the interaction energy must become positive at very short distances due to the Born repulsion—not considered in the original DLVO theory—preventing the interpenetration of proteins and NP (Adamczyk and Weronski, 1999).

The resulting potential interaction between NP and Tf is described by Equation 2 as a function of the distance $d$ between the center of mass of the protein and the closest point of the NP surface, as

$$U_{DLVO}(d) \equiv U_{VdW}(d) + U_{Elec}(d), \quad (2)$$

with Equation 3

---

1 We checked that in mono-component protein solutions of Tf our results have no qualitative dependence on small variation of the NP radius by preliminary calculations for $R_{NP}$ = 41 nm (Jareño and Delia, 2015).

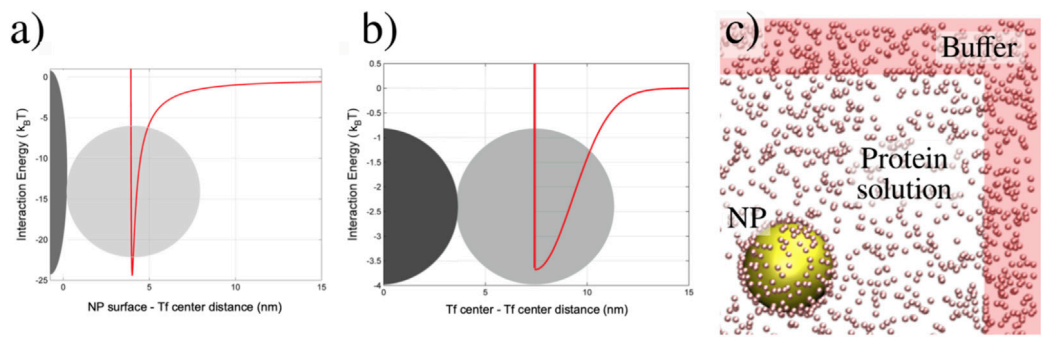

FIGURE 2
Schematic definition of the model. **(A)** The continuous line represents the protein–NP interaction potential (Equation 2) as a function of the distance between the protein center and the NP surface. The protein is represented as a sphere (light gray) with the Tf hydrodynamics radius. Only a small portion of the NP surface (dark gray) is sketched as a reference. **(B)** Protein-protein 3-body interaction potential (continuous line, Equation 5 for the case in which one protein (dark gray) is adsorbed onto the NP surface (not shown) and another protein (light gray) is approaching the first from the solution, as a function of the center of mass distance between the two proteins. **(C)** Snapshot of a portion of the simulation box showing the reaction region (white background) with the NP (large yellow sphere) in the center and proteins (small pink spheres) in the solution. The buffer region (red background) encompasses the reaction region.

$$U_{VdW}(d) \equiv \frac{1}{d^7} \frac{A_H \sigma^6}{2520} \frac{2R_{Tf} R_{NP}}{R_{Tf} + R_{NP}} - \frac{1}{d} \frac{A_H}{12} \frac{2R_{Tf} R_{NP}}{R_{Tf} + R_{NP}} \quad (3)$$

and Equation 4

$$U_{Elec}(d) \equiv \frac{64\pi k_B T \gamma_{Tf} \gamma_{NP} \rho_\infty}{\xi^2} \frac{R_{Tf} R_{NP}}{R_{Tf} + R_{NP}} e^{-\xi d} \quad (4)$$

where $\xi = 0.165 nm^{-1}$ is the inverse of the Debye-Huckel screening length, $\gamma_{Tf}$ and $\gamma_{NP}$ are the reduced surface potential of Tf and the NP, respectively, defined as $\gamma \equiv \tanh[ze\phi/4k_BT]$, $\phi$ is the corresponding zeta potential, $ze = 1\,e$ the valence of ions in solution and $\rho_\infty$ their concentration in bulk, $\sigma = 0.5$ nm is the characteristic parameter of the Born repulsive term; $A_H = 15 k_B T$ is the Hamaker constant. This value has been chosen to create a strongly bound layer on top of the surface while having a negligible effect on proteins far from it. The resulting interaction potential has a single minimum with energy of $-16.5\,k_BT$ (Figure 2A). Our model is calibrated using experimental parameters, with the effect of ions incorporated by calculating the protein binding affinities to nanoparticles directly from data obtained under specific experimental conditions. As a result, any changes in ionic strength or other experimental factors, such as protein charge, will affect the protein's affinity for the nanoparticles, requiring a recalibration of the model's phenomenological parameters.

Once a complete layer of protein forms on the NP surface, other proteins in the solution cannot interact directly with it due to the limited range of the NP-Tf interaction (Equation 2). However, experimental results suggest that proteins near the NP surface can still interact by forming new layers (Milani et al., 2012). While not all the proteins in the first layer are necessarily part of the HC, characterized by almost irreversible protein adsorption, those in the subsequent layers are typically weakly adsorbed and form the SC (Sharma et al., 2024).

We hypothesize that the aggregation of Tf occurs because of conformational changes in the proteins that are adsorbed within the HC. The rationale behind this hypothesis lies in the observation that the Tf is not prone to aggregation when it is suspended in solution under the specific thermodynamic conditions we are considering. This is consistent with the repulsive protein-protein interaction we use in Equation 1. Therefore, the emerging attractive force between free Tf and HC proteins must be due to the effects on the HC proteins caused by the surface of the NPs.

Specifically, when Tf is folded in solution, its hydrophobic amino acids are mainly shielded from water, as occurs for other structured mesophilic proteins (Bianco et al., 2017). Still, Tf can undergo partial unfolding when tightly bound to the NP, as it has been reported for many other proteins forming a corona (Park, 2020), exposing residues with a significant affinity toward other Tf proteins in the solution. As a result, proteins on the surface are highly likely to attract other proteins to minimize the overall free energy of the system. We assume that this process depends on how strongly each Tf binds to the NP via the Eq (ef DLVO), i.e., it depends on the Tf-NP surface distance.

Hence, we describe this attractive interaction between two Tf proteins $i$ and $j$ in Equation 5 as a 3-body potential that depends on their distances from the NP surface, $d_i$ and $d_j$ respectively, and their relative distance $r_{ij}$

$$U_{3b}(r_{ij}, d_i, d_j) \equiv -\varepsilon_{3b} \exp\left[-\frac{d_i d_j}{\kappa^2}\right] \exp\left[\frac{(r_{ij} - \delta)^2}{2\omega^2}\right] \quad (5)$$

where the characteristic interaction energy between proteins, $\varepsilon_{3b}$, is modulated by a decreasing exponential term depending on both $d_i$ and $d_j$, being maximum when one of the two proteins is adsorbed onto the NP surface. The modulation is controlled by the parameter $\kappa$. As discussed in the following, we must extrapolate our calculations to low values of $C_{NP}$ to compare our simulation results with the experimental data. However, as mentioned above, the phenomenological model parameters depend on the thermodynamic condition, specifically, the NP concentration $C_{NP}$. Therefore, for each $C_{NP}$, we adjust the model's parameter and find (as shown in the Results section) that, by fixing $\varepsilon_{3b} = $ $3.75 k_B T$ and varying $\kappa$, we can correctly match our extrapolations to the experimental findings.

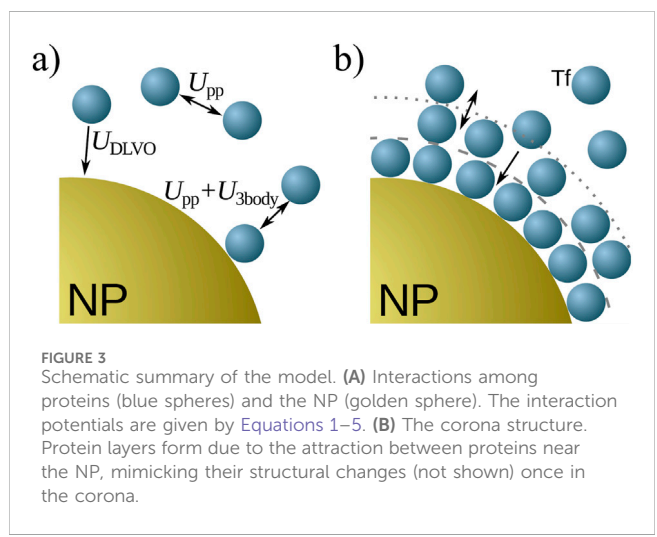

FIGURE 3
Schematic summary of the model. **(A)** Interactions among proteins (blue spheres) and the NP (golden sphere). The interaction potentials are given by Equations 1–5. **(B)** The corona structure. Protein layers form due to the attraction between proteins near the NP, mimicking their structural changes (not shown) once in the corona.

The second term in Equation 5 is a local attractive Gaussian well potential, centered at $\delta = 2R_{Tf}$ with width $\omega = \delta/4$, depending on the relative protein distances (Figure 2B). The 3-body interaction $U_{3b}$ sums up to the pair interaction $U_{pp}$ between proteins (Figure 3).

## 2.3 Computational details

The code we use here is available online as the simulation package BUBBLES (Vilanova, 2015) and was initially introduced by Vilanova et al. in Vilanova et al. (2016). Within this work, we provide a user-friendly tutorial, available on the reference website to use BUBBLES to study HC and SC formation for NPs of different chemistry in contact with solutions including one or more proteins.

We simulate systems at different Tf concentrations, $C_{Tf}$, to identify the parameters best reproducing the experimental data. In particular, we tune the repulsion energy between two adsorbed proteins at the shorter diameter distance, the zeta potential in PBS, and the Hamaker constant for the model potential with the NPs, Equation 2. BUBBLES controls $C_{Tf}$ implementing a protein reservoir (buffer) that we describe below.

The NP concentration remains constant at $C_{NP}$ = 0.1 mg/mL in the experiments. This value would require extraordinarily long simulations to equilibrate due to the low amount of proteins in solution, especially under extreme conditions of low [Tf]/[NP]. To address this issue, we maintain a high enough NP concentration $C_{NP}$ to ensure reasonable simulation times. Later, we vary $C_{NP}$ and adjust the relevant parameters at different NP concentrations. Finally, we extrapolate our results to the experimental NP concentration $C_{NP}$ = 0.1 mg/mL.

### 2.3.1 Langevin dynamics

We use Langevin dynamics to simulate protein diffusion. Langevin dynamics extend the Newtonian equations of motion to account for the effects of a surrounding molecular environment, such as particles in a solvent, with an implicit description (Schneider and Stoll, 1978). The set of equations of motion of a given protein $i$ are described by Equation 6:

$$\begin{aligned} \vec{v}_i(t) &= \frac{d\vec{r}_i(t)}{dt} \\ m_i \vec{a}_i(t) &= \vec{F}_i(t) - \gamma m_i \vec{v}_i(t) + \vec{R}_i(t), \end{aligned} \quad (6)$$

where $m_i$ is the mass of the protein; $\vec{r}_i(t)$, $\vec{v}_i(t)$ and $\vec{a}_i(t)$ are the instantaneous coordinates, velocity and acceleration, respectively. Here Equation 7

$$\vec{F}_i(t) \equiv \sum_{j \neq i} \vec{f}_{ij}(t) \quad (7)$$

is the result of the interparticle forces described in Equation 8

$$\vec{f}_{ij} \equiv -\vec{\nabla}_i \left[ U_{pp}(r_{ij}) + U_{DLVO}(d_i) + U_{3b}(r_{ij}, d_i, d_j) \right] \quad (8)$$

acting on the protein $i$, where the interaction potentials are defined by the Equations 1–5.

Two additional terms appear in the inertial part of the system to account for the proteins' interaction with the solvent. The first is the viscous force $\gamma m_i \vec{v}_i(t)$ felt by a protein traveling through the solvent, being $\gamma$ the viscosity. The second is the stochastic force $\vec{R}_i(t)$ associated with the momentum transfer of the solvent molecules to the proteins through random collisions. This stochastic force is modeled as a Gaussian noise of zero mean and variance $\langle \Delta \vec{R}(t) \Delta \vec{R}(t') \rangle = 2 k_B T \gamma m \delta(t - t')$, where $T$ is the temperature of the solvent—acting as a thermostat for the system—and $\delta(t)$ is the Dirac delta function because the force is uncorrelated in time. The numerical integration in time of the equations is implemented via a modified version of the velocity-Verlet integrator, i.e., the Brünger-Brooks-Karplus (BBK) integrator (Brünger et al., 1984).

### 2.3.2 Buffer implementation

In the experiments, the concentration of NPs, $C_{NP}$ = 0.1 mg/mL, is up to four orders of magnitude smaller than the protein concentration, with tens of thousands of proteins per NP. Under these conditions, the interaction between the protein and NP occurs in a localized region much smaller than the overall system volume, and the simulation times are exceedingly long. To enhance the simulation performance, we narrow our focus to a specific system section centered on a single NP while treating the remaining proteins as a buffer to maintain constant protein concentration in the localized region. This approach reduces the computational cost of simulating the dynamics of a large set of particles (Brooks and Karplus, 1983; Brooks and Karplus, 1989).

The main idea is to use a particle reservoir to regulate protein concentration within the area of interest. A similar approach involves conducting simulations in the macrocanonical ensemble, using the chemical potential to control the concentration (Oberholzer et al., 1997). Our method does not require particle insertion/deletion events, which can be inefficient at high concentrations. Additionally, this method allows us to simulate a system with a constant number of particles, which is very convenient, especially for GPU-based numerical simulations, due to device memory management restrictions.

We divide the simulation box into two centered cubic regions: an inner region of volume $V_r$ called the *reaction region* and an outer region of volume $V_b$ called the *buffer region*. The reaction region contains the NP in its center, where all the protein adsorption processes occur. The buffer region surrounds the reaction region. The inner walls of the buffer overlap with the walls of the reaction region and are semipermeable, trapping excess proteins from the reaction region. The outer walls of the buffer region have periodic boundary conditions (Figure 2C).

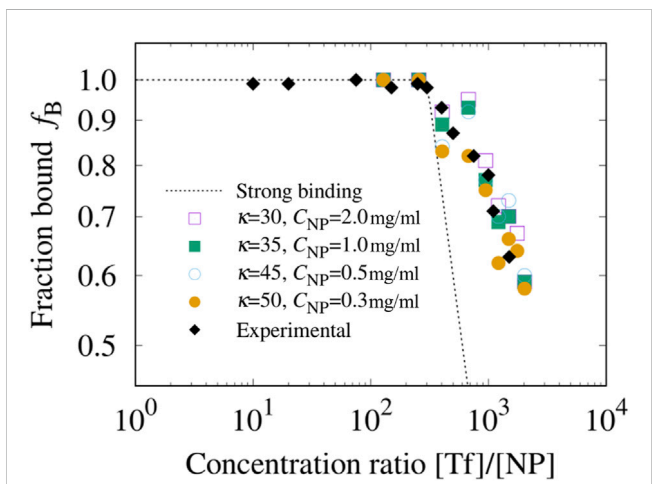

FIGURE 4
Experimental Tf fraction bound polystyrene NPs compared with the model's calculations. We use the experimental fraction bound (black diamonds) as a function of the relative concentration [Tf]/[NP] in Milani et al. (2012). to calibrate the model's parameter $\kappa$ in Equation 5 by best-fitting the calculated $f_B$ when we fix $\varepsilon_{rm3b} = 3.75 k_B T$ and vary $C_{NP}$ from 0.3 mg/mL (orange circles), to 0.5 mg/mL (white circles), to 1.0 mg/mL (green squares) to 2.0 mg/mL (white squares). The dashed line shows how $f_B$ would change with [Tf]/[NP] if the proteins could only bind strongly to the NP's surface without forming a soft corona. The dashed line and the data depart at [Tf]/[NP] $\simeq 320$. We also notice a change in the slope of the data for [Tf]/[NP] between $\simeq 700$ and $\simeq 1000$.

During the simulation, we calculate the difference

$$\Delta C \equiv C - C_r \quad (9)$$

between the concentration $C_r \equiv n_r M_{Tf}/V_r$, of $n_r$ free proteins in the $V_r$ reaction region, and the concentration of the reference system $C$, as described by Equation 9. Proteins that adsorb into the NP's hard and soft corona do not contribute to $C_r$.

If $\Delta C > 0$, we re-equilibrate the protein concentration in $V_r$ by letting $n_+ \equiv \Delta C V_r/M_{Tf}$ proteins diffuse from the buffer to the reaction region, choosing them among all the $n_b$ proteins in $V_b$ with a probability $p_+ \equiv n_+/n_b$. On the other hand, if $\Delta C < 0$, we choose $n_- \equiv |\Delta C| V_r/M_{Tf}$ proteins in $V_r$, with probability $p_- \equiv n_-/n_r$, and let them diffuse. If they cross the semipermeable wall of the buffer region, they can no longer diffuse back to the reaction region.

## 3 Results and discussion

### 3.1 Fraction bound and model calibration

The experiments by Milani et al. were carried out at the NP concentration of 0.1 mg/mL and a [Tf]/[NP] of up to approximately 1,000 (Milani et al., 2012). Even with our coarse-grained approach, the low NP and high protein concentration combination made numerical simulations impractical. We perform simulations at various higher NP concentrations to address this issue and then extrapolate the results to match the experimental conditions.

The majority of the parameters of our model are given by the experimental setup and discussed in Section. 2.2. Those for Equations 2–4 are set from the adsorption isotherms of a Tf monolayer, as discussed in Vilanova et al. (2016). To evaluate the phenomenological parameters for the interaction relevant to the SC (Equation 5), we set the NP concentration $C_{NP}$ and find the corresponding $\kappa$ and $\varepsilon_{rm3b}$ that best fit the experimental data (Milani et al., 2012) for the Tf fraction bound

$$f_B \equiv \frac{N_{Ads}}{N_{Tot}} \quad (10)$$

where $N_{Ads}$ is the number of adsorbed proteins, and $N_{Tot}$ is the total number of proteins in the system as a function of the relative concentration [Tf]/[NP], where [Tf] $\equiv N_{Tot}/V$ and [NP] $\equiv N_{NP}/V$ are the number densities of Tf and NPs, respectively, in the system's volume $V \equiv V_r + V_b$ (Figure 4). In our simulations, as discussed in Section. 2.3, we set $N_{NP} = 1$.

As long as $N_{Tot} < N_{max}$, the maximum number of Tf that can be adsorbed onto the NP, $f_B$ (Equation 10) follows the *strong binding* prediction in which all the proteins in solution end up into the corona. For a polystyrene NP with radius $R_{NP} = 35$ nm, the strong binding is observed for [Tf]/[NP] $\leq 320$, consistent with our numerical estimate for $N_{max} = 320$. We find $N_{max}$ by simulating a supersaturated protein solution.

In the absence of a soft corona, for [Tf]/[NP] $> 320$, the proteins in excess would not be adsorbed onto the NP's surface. Then, it would be $f_B = N_{max}/(N_{max} + N_{exc})$, that for $N_{max} \gg N_{exc}$ is a rapidly decreasing function of [Tf]/[NP] = $N_{max} + N_{exc}$, $f_B \simeq 1 - c \ln([Tf]/[NP])$, where $c = 1/[(\ln N_{max}) N_{max}/([Tf]/[NP] - N_{max})]$ is a large number approximately independent of [Tf]/[NP][2]. Instead, at any [Tf]/[NP] > 320, the experimental $f_B$ is larger than the strong-binding prediction (Figure 4), demonstrating that Tf forms a soft corona, as discussed by Milani et al. (2012).

Our model reproduces this behavior associated with the formation of the SC. Once a layer is formed, the next layer can only form at a greater distance from the NP surface. This means that the interaction between proteins, decreases as their average distance from the surface increases. As a result, the new layer will attract fewer proteins, leading to a decrease in the slope of $f_B$ as a function of [Tf]/[NP]. Consistently, we observe a sudden decrease in the slope when [Tf]/[NP] $\simeq 320$, indicating the saturation of the first layer. This effect becomes weaker for subsequent layers, with a minor change in slope observed at [Tf]/[NP] between $\simeq 700$ and $\simeq 1000$, corresponding to the saturation of the second layer and the formation of the third in the protein corona, as discussed in Section 3.2.

We find that for $C_{NP}$ ranging from 0.3 mg/mL to 2.0 mg/mL, the best fit of the experimental data occurs when $\varepsilon_{3b} = 3.75 k_B T$, regardless of $C_{NP}$, and $\kappa$ decreases monotonically and regularly as $C_{NP}$ increases (Figure 4). Since an increase in $C_{NP}$ implies an increase in [Tf] at a fixed [Tf]/[NP], this finding shows that as the protein concentration rises, the interaction range of the three-body potential in Equation 5 needed to match the experimental data decreases. This result aligns with the idea that recruiting proteins from a greater distance is necessary to populate the SC at a lower protein density.

---

2 For $N_{max} \gg N_{exc}$, $\ln([Tf]/[NP]) \simeq \ln N_{max} + N_{exc}/N_{max}$ and $f_B \simeq 1 - N_{exc}/N_{max} \simeq 1 - \ln([Tf]/[NP])(1 - \ln N_{max}/\ln([Tf]/[NP])) \simeq 1 - c \ln([Tf]/[NP])$

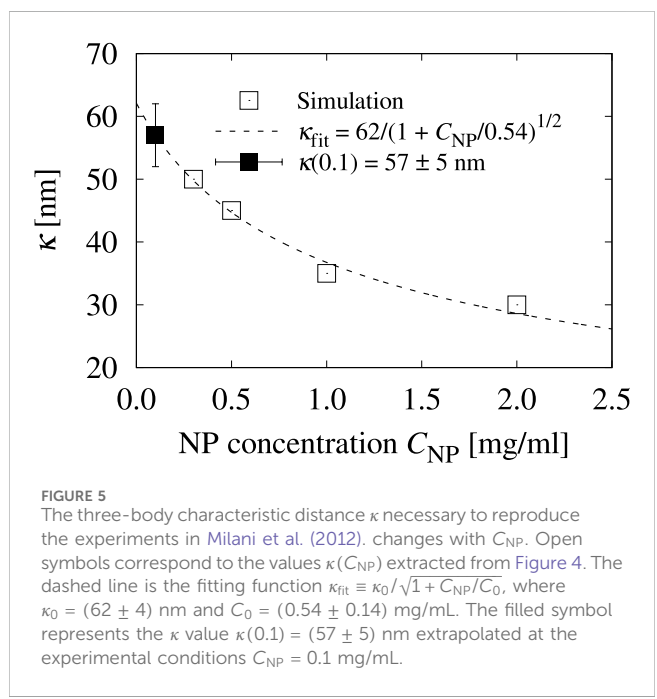

FIGURE 5
The three-body characteristic distance $\kappa$ necessary to reproduce the experiments in Milani et al. (2012). changes with $C_{NP}$. Open symbols correspond to the values $\kappa(C_{NP})$ extracted from Figure 4. The dashed line is the fitting function $\kappa_{fit} \equiv \kappa_0/\sqrt{1 + C_{NP}/C_0}$, where $\kappa_0 = (62 \pm 4)$ nm and $C_0 = (0.54 \pm 0.14)$ mg/mL. The filled symbol represents the $\kappa$ value $\kappa(0.1) = (57 \pm 5)$ nm extrapolated at the experimental conditions $C_{NP} = 0.1$ mg/mL.

We observe that the estimates for $\kappa$ follow a regular function of $C_{NP}$ that is well represented by Equation 11

$$\kappa = \frac{\kappa_0}{\sqrt{1 + C_{NP}/C_0}}, \qquad (11)$$

where $\kappa_0$ and $C_0$ are constants (Figure 5). By performing the least squares fitting of $\kappa$, we obtain that $\kappa_0 = (62 \pm 4)$ nm and $C_0 = (0.54 \pm 0.14)$ mg/mL. This expression can be easily rationalized as $(\kappa/\kappa_0)^2 \sim 1/(1 + C/C_0)$ implies that the surface, proportional to $\kappa^2$, at which the SC recruits proteins at fixed fraction bound decreases linearly with the protein concentration in the low-$C_{NP}$ limit, i.e., $\kappa^2 \sim \kappa_0^2 - [Tf]/[Tf_0]$, where $[Tf_0] \equiv [Tf]C_0/C_{NP}$. Extrapolating $\kappa$ at $C_{NP} = 0.1$ mg/mL, we estimate $\kappa = (57 \pm 5)$ nm as the appropriate value for the experimental conditions in Milani et al. (2012), Figure 5.

It is important to note that $\kappa$ does not control the interaction range of Equation 5. The three-body interaction range is actually determined by the width $\omega = \delta/4 = R_{Tf}/2$ of the Gaussian centered at $\delta = 2R_{Tf}$, which is small compared to the protein size, $\omega/(2R_{Tf}) = 0.25$.

On the other hand, the parameter $\kappa$ regulates how far the NP can affect the proteins, inducing an effective attraction among them, and should be compared with the extension of the SC. As discussed below, the SC can reach $R_{SC} \simeq 20$ nm. Hence, it is $\kappa/R_{SC} \simeq 2.8$, suggesting that, within our model, the corona can, on average, induce protein structural fluctuations at more than twice its size under the experimental conditions in Milani et al. (2012).

## 3.2 Protein corona density profile

First, we examine the structure of the protein corona in simulated solutions at relative concentrations [Tf]/[NP] > 320, at which data in Figure 4 suggest the formation of the SC. As discussed in the previous section, we set $C_{NP} = 1$ mg/mL and $\kappa = 35$ nm to ensure the feasibility of the simulations.

We calculate the local density of proteins within distances $r$ and $r + \delta r$ with $\Delta r = 0.1$ nm, i.e., the protein density profile as a function of the proteins' distance $r$ from the NP surface. We then normalize the density profile by the average protein concentration $C_{Tf}$, resulting in the protein radial distribution function (RDF), defined in Equation 12 as

$$g(r) \equiv \frac{1}{[Tf]V(r, r + \Delta r)} \langle \sum_{i}^{N_{Tf}} \delta(r_i - r_{NP} - r) \rangle \qquad (12)$$

where $V(r, r + \Delta r) \equiv (4/3)\pi[(r + \Delta r)^3 - r^3]$ is the spherical shell at distance $r$ and thickness $\Delta r$, $N_{Tf}$ is the total number of Tf proteins in solution and $\delta(x)$ is the Dirac delta function. We denote $\langle \cdot \rangle$ as the ensemble average over uncorrelated configurations (Figure 6).

In this range of [Tf]/[NP] values, the RDF displays up to 3 peaks corresponding to the different protein corona layers. The first peak is centered at $r_1 \simeq R_{Tf} = 3.72$ nm and is independent of the relative concentration for [Tf]/[NP] ≥ 400. This is consistent with the observation that the first layer saturates for [Tf]/[NP] = 320 (Figure 4). The sharpness of the peak indicates strong protein adsorption, as expected in the HC.

The second layer, at approximately $r_2 \simeq 3R_{Tf} = 11.16$ nm, is well separated from the first and is broader. It displays a pre-peak between $\simeq 2R_{Tf}$ and $\simeq 2.5R_{Tf}$, followed by a prominent peak extending up to $\simeq 3.5R_{Tf}$. These ranges are consistent with soft interactions of the proteins within the first two layers, as expected for the SC. The second layer is populated for all the concentrations considered here, and its intensity increases at larger [Tf]/[NP], approaching saturation at large [Tf]/[NP]. This is consistent with the change of slope observed in Figure 4 at [Tf]/[NP] > 700, as expected for the formation of a complete second layer. We label this layer as $SC_1$.

The third peak, centered around $r_3 \simeq 17.5$ nm $< 5R_{Tf} = 18.6$ nm, merges with the second layer and expands up to $\simeq 5.5R_{Tf}$ decaying within the protein solution at longer distances. It is broader than the other two peaks, suggesting a softer interaction with the NP than the second layer. This peak only appears for the two highest relative concentrations considered here, with [Tf]/[NP] ≤ 1000, when the second peak has reached its saturation level. We refer to this layer as $SC_2$.

We do not observe a saturation of the $SC_2$ layer within the investigated concentration ranges nor detect any distinct peaks above 20 nm. Instead, the density profile $g(r)$ approaches unity, as expected for a homogeneous solution. These observations suggest that the proteins in the $SC_2$ layer freely exchange with those in suspension.

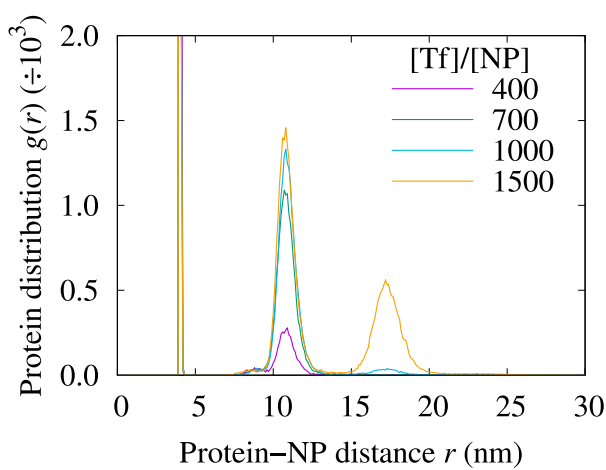

FIGURE 6
The density profiles $g(r)$ of the Tf proteins adsorbed within the corona, as a function of their distance $r$ from the polystyrene NP surface, calculated by our simulations and divided by a factor $10^3$. The three separate peaks correspond to three different layers of proteins. The layers broaden as they are farther from the NP surface, with the outermost occupied only for the two largest values of [Tf]/[NP]. At distances larger than the third layer, the local concentration converges toward the average value within the available volume, corresponding to $g(r) \simeq 1$. Lines correspond to four different [Tf]/[NP] ratios at which we expect the SC formation based on Figure 4: 400 (indigo), 700 (green), 1,000 (blue), 1,500 (orange). All data are at $C_{NP}$ = 1 mg/mL, setting $\kappa$ = 35 nm.

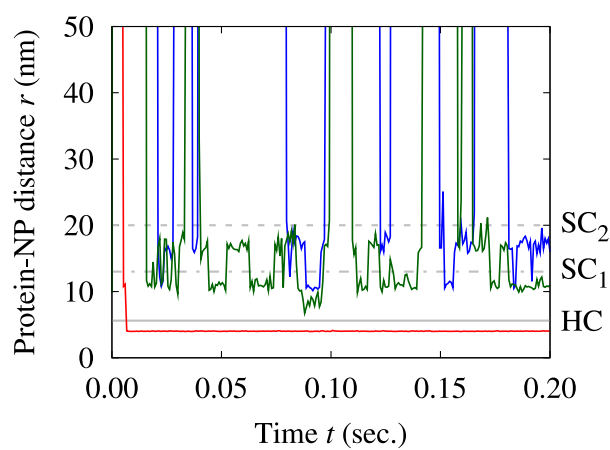

FIGURE 7
Evolution of the distance from the NP surface of three single proteins over time. We select three proteins among those that spend most of their time within the first (red track), the second (green track), and the third layer (blue track). The gray lines mark the largest distances for each layer, as defined in Figure 6: continuous at 5.6 nm for HC, dotted-dashed at 13 nm for $SC_1$, and dashed at 20 nm for $SC_2$. The protein adsorbed within the HC layer remains attached throughout the simulation, while those in the SC layers exchange frequently with the suspension. The simulation conditions are as in Figure 6.

## 3.3 Protein corona dynamics and glassy behavior

### 3.3.1 Irreversible and reversible adsorption

Next, we investigate the adsorption kinetics of Tf onto the NP. To validate the interpretation from the RDF analysis of the three layers as hard and soft corona, we track the positions of the proteins within them (Figure 7).

The tracking confirms the strong adsorption of proteins in the first layer, as they show no signs of displacement once adsorbed. On the other hand, proteins in the other two layers are constantly exchanged with the suspension. They can detach from the protein corona, return to the protein solution, and eventually get reabsorbed, regardless of whether they are in the second or third layer. Our simulations suggest that the proteins in the second layer have longer residence times than those in the third layer, indicating higher stability of the inner layer of the soft corona than the outer layer.

Proteins in the outermost layer sometimes interchange with those in the second layer, but we do not find exchanges of any proteins in the two outer layers with the first layer. All these observations consistently associate the innermost layer with the HC and the two outermost layers with the SC, validating the conclusions drawn from the RDF analysis.

Further understanding can be reached by analyzing the time evolution of the number $N_{Ads}$ of proteins adsorbed within each corona layer (Figure 8). In the first layer (Figure 8A), $N_{Ads}$ saturates very rapidly at a value that depends only weakly on [Tf]/[NP] and is consistent with the experimental estimate of [Tf]/[NP] $\simeq$ 320 (Milani et al., 2012) (Figure 4). The higher [Tf]/[NP], the shorter the time needed to reach saturation of the first layer, with saturation time estimates ranging from 0.1 to 1 s for the concentrations considered here. Once saturated, $N_{Ads}$ shows only minor fluctuations within the first layer, as expected for strongly bound proteins within the HC.

In the second layer (Figure 8B), $N_{Ads}$ increases with [Tf]/[NP] and approaches saturation for [Tf]/[NP] > 675, consistent with the change of slope observed in the experimental data for the fraction bound (Figure 4). The layer exhibits reversible binding as emphasized by the fluctuations $N_{Ads}$ over time. The intensity of the fluctuations decreases when the $SC_2$ layer approaches the saturation.

These fluctuations increase, instead, in the third layer (Figure 8C), being consistently higher than in the second layer, indicating that the corona gains stability closer to the NP surface. The $SC_3$ layer, populated only for [Tf]/[NP] ≤ 675, does not saturate for the concentrations considered here and always shows reversible absorption, as expected for the SC.

### 3.3.2 The soft corona glassy behavior

To better characterize the soft corona dynamics, we analyze the autocorrelation function with Equation 13 for the proteins populating the $SC_1$ and $SC_2$ layers,

$$C_i(t) \equiv \frac{\langle N_{Ads,i}(t_0) N_{Ads,i}(t_0 + t) \rangle - \langle N_{Ads,i}(t_0) \rangle \langle N_{Ads,i}(t_0 + t) \rangle}{\langle N_{Ads,i}(t_0)^2 \rangle - \langle N_{Ads,i}(t_0) \rangle^2}$$

(13)

where $N_{Ads,i}(t)$ is the time-dependent number of Tf proteins within the layer $SC_i$ with $i$ = 1, 2, and the averages are taken over the initial time $t_0$ larger than the characteristic time needed for the layers to stabilize its population, i.e., $\simeq$2 sec for [Tf]/[NP] = 1,500 (Figures 8B, C).

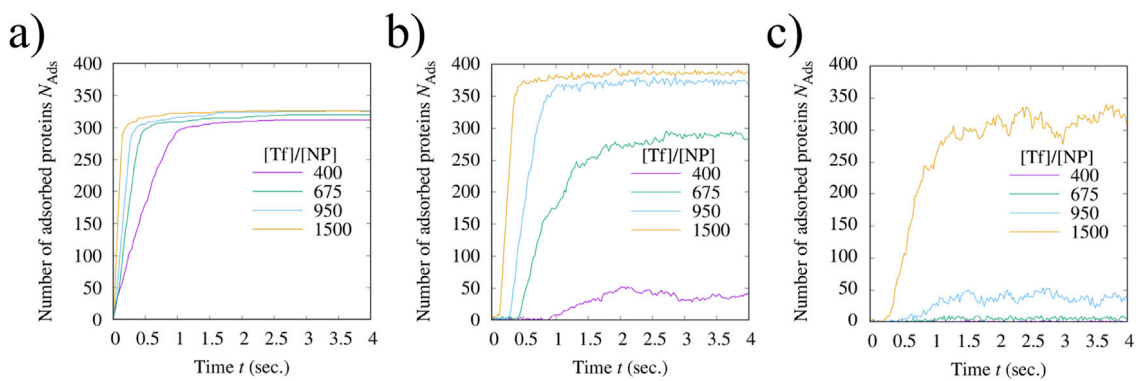

FIGURE 8
Simulation results for the number of Tf proteins adsorbed into the corona of a polystyrene NP over time at four different relative concentrations [Tf]/[NP]. The time evolution of this number varies among **(A)** the first, **(B)** the second, and **(C)** the third layer, with larger fluctuations for the outer layers, consistent with our interpretation of the first as the HC and the others as the SC. In each panel, the considered values of [Tf]/[NP] are 400 (indigo lines), 675 (green lines), 950 (turquoise lines), and 1,500 (orange lines). The simulation conditions are as in Figure 6.

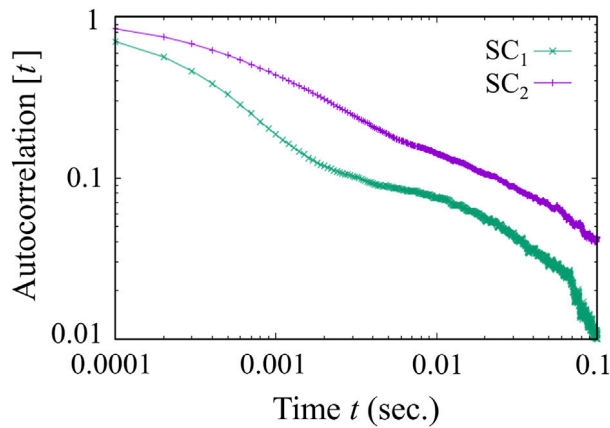

FIGURE 9
Density autocorrelation function of the soft corona over time. While the outer layer $SC_2$ has an approximate power-law decay (linear in double-logarithmic scale) over two decades, the inner layer $SC_1$ deviates from it, displaying a glassy dynamics. The simulation conditions are as in Figure 6 for [Tf]/[NP] = 1,500.

The two SC layers exhibit distinct behaviors, as shown in Figure 9. The $C_2(t)$ for the outer layer $SC_2$ demonstrates a decay pattern that departs from exponential, appearing closer to a power law across two decades. Exponential behavior would be anticipated if the system had achieved equilibrium, resulting in decorrelated population fluctuations of $SC_2$ over a characteristic time scale. A power-law decay suggests that such a time scale does not exist, indicating that the fluctuations remain correlated over durations exceeding our observation window of 0.1 s.

Conversely, the inner layer of the soft corona, $SC_1$, begins with a rapid decay, transitioning into a plateau that persists for nearly one decade. Over longer timeframes, $C_1(t)$ departs from the plateau, potentially adopting a power-law decay. This non-exponential behavior in the correlation function is characteristic of glassy systems, where dynamic progress is confined within local free-energy minima. The greater the system's distance from equilibrium, the higher the plateau in the autocorrelation function (Gotze and Sjogren, 1992; Kumar et al., 2006).

The relaxation behavior in the SC's inner layer implies that the corona's dynamic evolution in this volume is affected by factors such as crowding from other proteins within the corona. These factors can restrict protein mobility, leading to the development of dynamically *frozen* protein clusters or aggregates, which contribute to the relaxation behavior observed.

## 3.4 Model validation

To experimentally validate our *in silico* model for the protein soft corona simulation, we use DCS to evaluate the size distribution of polystyrene NPs before and after exposure to Tf at varying concentrations. Unlike the reference experiment for our modeling, (Milani et al. (2012)), our 100 nm PSCOOH polystyrene NPs exhibit an effective radius that closely matches the nominal radius. Specifically, in pristine conditions, our DCS data reveal a sharp, monodisperse distribution of NPs with an average diameter of ≃110 nm (Figure 10). The DLS results support this observation, with a z-average of 108.7 ± 1.2 nm and a polydispersity index (PDI) of 0.018 ± 0.005, as well as a zeta potential of −65.7 ± 2.4 mV. This diameter exceeds the one used in the simulations based on Milani et al. (2012). Consequently, when we set the NP concentration to match that of the simulations, i.e., $C_{NP}$ = 1 mg/mL, the corresponding [NP] is lower than in the simulations. To address this in the experiments, we adjusted the protein concentration to maintain the same [Tf]/[NP] ratios of 400, 675, 950, and 1,500 as in the simulations. At any [Tf]/[NP] > 0, DCS displays a main peak marking the apparent diameter of the NP-corona complex. The primary peak shifts to higher diameter values at higher [Tf]/[NP], from ≃ 130 nm at [Tf]/[NP] = 400 to ≃ 150 nm at [Tf]/[NP] = 1500 (Table 1).

We observe only one peak in the DCS measurements at [Tf]/[NP] = 1500, indicating that the NP-corona complexes do not aggregate under these conditions. However, at lower [Tf]/[NP], we find a structured distribution of apparent diameters corresponding to aggregates of different sizes. This observation suggests that aggregation is inhibited under crowded conditions. Nevertheless, the increase in the apparent diameter of the NP-corona complex (Table 1) emphasizes the increase in the corona's size as [Tf]/[NP] increases, consistent with our simulation results.

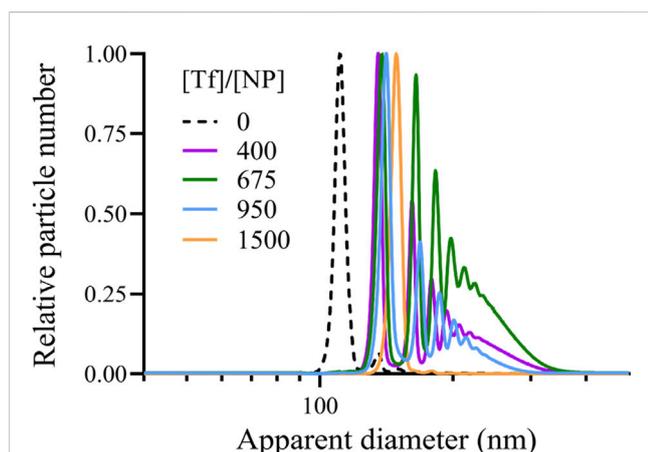

FIGURE 10
DCS measurements of the (relative) number of polystyrene NPs as a function of their apparent diameter after exposure to Tf proteins at relative concentration [Tf]/[NP]. We find one peak at large [Tf]/[NP] marking the apparent diameter and a main peak—corresponding to a single NP-corona complex diameter—followed by secondary maxima at lower [Tf]/[NP] due to a larger tendency to form clusters under these diluted conditions. [Tf]/[NP] goes from 0 (dashed line) to 400 (indigo line), 675 (green line), to 950 (blue line) to 1,500 (yellow line).

To make a quantitative comparison of the simulation predictions with the experimental measurements, we estimate for each condition the average number $N_{Ads}$ of Tf proteins adsorbed onto each NP across the various layers of the protein corona. To this end, we apply a core-shell model to the apparent size values marked by the main DCS peaks. We determine the thickness of the protein layer adsorbed onto the NPs and subsequently derive the number of proteins adsorbed (Table 1).

The experiments show a satisfactory quantitative agreement with the predictions based on the simulations of our coarse-grained model (Figure 11). We find that the model slightly underestimates or overestimates $N_{Ads}$ at small and large values of [Tf]/[NP], with $\simeq -33\%$ and $\simeq 18\%$ deviations at [Tf]/[NP] = 400 and 1,500, respectively, while agrees within the standard fluctuations at intermediate values, with minor deviations of $\simeq 1.6\%$ and $\simeq 7\%$ at [Tf]/[NP] = 675 and 950, respectively. Therefore, the model simulations are in semi-quantitative agreement with the experiments, validating our approach and coarse-grained model for the hard and soft corona.

## 4 Discussion, conclusion, and perspectives

When NPs come into contact with biological fluids, biomolecules adhere to their surfaces, forming a corona composed of multiple layers that influence how these NPs interact with cells and their biological effects. In protein-rich environments, such as blood or serum, strong protein-NP interactions form a hard corona. In contrast, weaker protein-protein and protein-NP interactions give rise to a soft corona. The hard corona is characterized by proteins that adsorb irreversibly onto the NP, whereas the soft corona consists of a fluid layer of loosely bound proteins associated with the NP. Understanding the dynamics of each component of the protein corona is essential for various biological applications of nanotechnology (Sharma et al., 2024).

However, measuring the soft corona poses significant experimental challenges (Guo et al., 2024). *In-situ* detection methods, which aim to preserve the protein corona in its physiological environment, demand advanced technical skills and involve complex procedures, such as tagging proteins with fluorescent markers and using dynamic light scattering. These modifications risk altering protein binding capacities, thereby complicating accurate measurements. *Ex-situ* methods, such as ultraviolet-visible spectroscopy and liquid chromatography-mass spectrometry, struggle with the incomplete separation of NP-protein complexes, which can lead to misidentifying the protein corona constituents. Additionally, excessive centrifugal force during separation may result in the loss of protein corona components, further complicating accurate detection.

Therefore, the composition of the protein corona has traditionally been examined, either *in vivo* or *in vitro*, through static incubation methods. Nevertheless, this approach limits our ability to explore the dynamics of the corona components over biologically relevant timescales. Here, we present new findings for a computational method that enables us to integrate experimental results and comprehensively describe the dynamics of both the hard and soft corona.

We extend the modeling and computational approach we previously introduced, incorporating the study of the soft corona

TABLE 1 Experimental analysis from DCS measurements.

| [Tf]/[NP] | DCS main peak (nm) | Corona thickness (nm) | $N_{Ads}$ |
|---|---|---|---|
| 0 | 110.1 | 0 | 0 |
| 400 | 135.6 | 2.9 | 550 |
| 675 | 138.2 | 3.2 | 615 |
| 950 | 141.4 | 3.6 | 700 |
| 1,500 | 149.0 | 4.7 | 915 |

We indicate for each [Tf]/[NP] > 0 the main peak position corresponding to an apparent diameter in nm for the NP-corona complex, the thickness in nm of the protein layer adsorbed onto the NPs, and the average number $N_{Ads}$ of proteins adsorbed within the corona per NP.

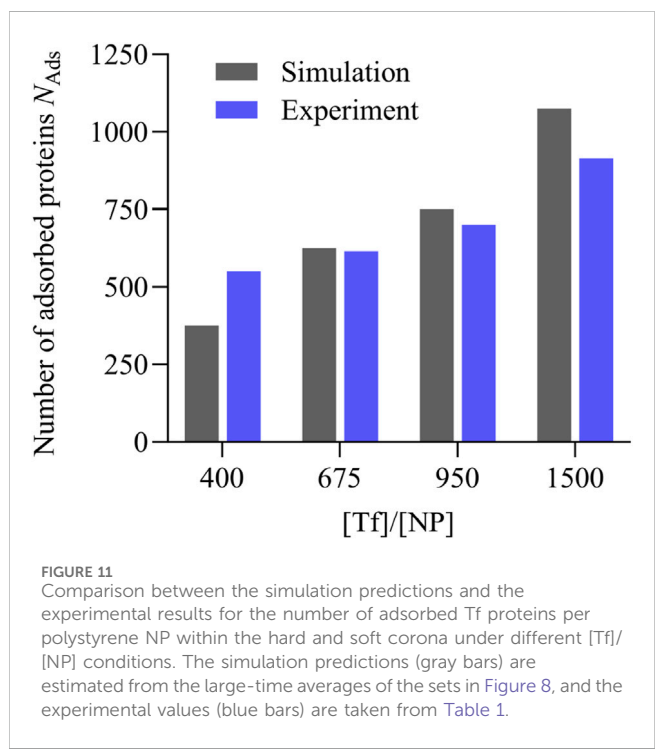

FIGURE 11
Comparison between the simulation predictions and the experimental results for the number of adsorbed Tf proteins per polystyrene NP within the hard and soft corona under different [Tf]/[NP] conditions. The simulation predictions (gray bars) are estimated from the large-time averages of the sets in Figure 8, and the experimental values (blue bars) are taken from Table 1.

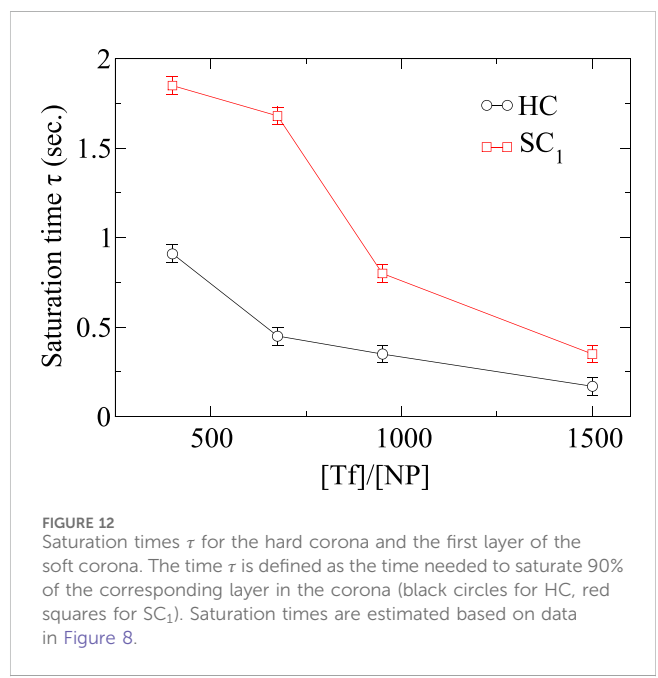

FIGURE 12
Saturation times $\tau$ for the hard corona and the first layer of the soft corona. The time $\tau$ is defined as the time needed to saturate 90% of the corresponding layer in the corona (black circles for HC, red squares for $SC_1$). Saturation times are estimated based on data in Figure 8.

into analyzing the short- and long-term kinetics of the hard corona formed by different proteins competing for the NP surface (Vilanova et al., 2016). The method employs a coarse-grained description of globular proteins interacting with the NP through colloid-like effective potential energies. This approach allows for molecular simulations over time scales on the order of seconds within the framework of Langevin dynamics once a few phenomenological parameters are extracted from preliminary experiments with monocomponent protein solutions. These simulation predictions can then be extended to any time scale and validated against laboratory results, utilizing a non-Langmuir adsorption theory introduced in Vilanova et al. (2016).

The extension introduced here incorporates a three-body interaction between proteins and the NP, mimicking the denaturation effect that the nanosurface can induce on proteins (Park, 2020). We assume that this effect exponentially decays as the distance between the NP's surface and any interacting proteins increases. The alignment of our simulation results with experimental data validates our model a posteriori.

As a case study, we consider the experimental conditions described in Milani et al. (2012), which evaluated the formation of the hard and soft corona by Tf proteins in suspension with polystyrene NPs. The experimental setup presents challenges due to the highly diluted concentrations of NPs and proteins, necessitating macroscopically large systems that are computationally prohibitive to simulate. To address this, we develop a scaling strategy to replicate the overall experimental behavior in smaller systems with higher concentrations and effective parameters. Using as an example the parameter describing the exponential decay of the three-body protein-protein-NP interaction, we show how this strategy allows us to extrapolate the quantity of interest in the limit of the experimental conditions.

After establishing the model, we used it to investigate the structure and kinetics of the corona. Our structural analysis reveals that the corona is composed of three distinct layers. The layer closest to the NP's surface is the sharpest, while the outermost layer is the broadest. The first and second layers become populated at low and moderate protein concentrations, whereas the third layer forms only at the highest concentrations.

The protein dynamics in each layer reveals that the first layer, which is in direct contact with the NP, contains proteins tightly bound to the surface. These proteins adsorb irreversibly over the simulation time scales, forming the hard corona. In contrast, the proteins in the second and third layers are reversibly bound to the corona, constituting the soft corona.

Our study shows that the first (HC) and second ($SC_1$) layers reach saturation at the protein concentrations examined and that, overall, the entire corona needs approximately 2 s to reach an apparent stable number of adsorbed proteins at the highest [Tf]/[NP] considered here. Furthermore, the time required to get 90% saturation within a given layer decreases as protein concentration increases, with the HC approaching a characteristic value of about 0.2 s. This trend is similar for the soft corona $SC_1$ layer, but the characteristic time to reach 90% saturation is about twice that for the HC (Figure 12). Although the protein concentrations are insufficient to saturate the third layer ($SC_2$), a similar trend is observed for the outermost part of the soft corona, with saturation times for $SC_2$ that are two or three times larger than $SC_1$. Overall, the time required to stabilize a layer at a given relative concentration [Tf]/[NP] increases from the hard corona to the outermost layer of the soft corona, ranging from approximately 0.2 s to around 1.0 s at the highest concentration considered in this study.

The in-depth analysis of the dynamics of the soft corona reveals that $SC_1$ displays a glassy behavior. Its density autocorrelation function has a nonexponential decay and develops a plateau at intermediate times. We understand this behavior as a consequence of the crowding of the inner layer caused by the outer one. Hence, although the time $\tau$ needed to reach 90% of its saturation for the soft corona is on the order of seconds, the corona could evolve over longer times as in glassy systems. We expect that at higher biomolecular concentrations, such as those in blood, this slowing down could become relevant for biological processes and should be taken into account when analyzing the effects of the protein soft-corona in the NP-cell interaction over time.

We validate our model predictions by directly estimating, under similar experimental conditions, the number of Tf proteins adsorbed onto 100 nm PSCOOH polystyrene NPs. Despite the differences, mainly related to the nominal size of the NPs, our experiments show that the computational model can predict within a 10% error the number of adsorbed proteins at the different relative concentrations [Tf]/[NP] considered here. We observe that the model is, overall, predicting a faster-increasing number of proteins adsorbed with the increase of the [Tf]/[NP] ratio relative to the experiments. This difference is possibly due to the agglomeration of the NPs observed in the experiment at increasing [Tf]/[NP]. The clustering of more than one NP-corona complex could 1) decrease the surface area available for protein binding and 2) lead to an experimental underestimation in the number of adsorbed Tf due to the density value used for Tf.

Lastly, the computational model does not consider Tf structural changes upon adsorption. Instead, recent findings from cryo-TEM corona show a uniform corona layer rather than a packing of globular proteins, suggesting structural deformation of the protein structure (Sheibani et al., 2021). This deformation could correspond to the protein flattening out, which would reduce the number of adsorbed Tf.

The overall conclusion is that the close correlation between the experimental and computational models in the present case study demonstrates satisfactory performance in protein-rich environments, indicating the applicability of our approach to biologically relevant conditions. For this reason, to benefit the scientific community, we provide an open-source interactive tutorial with all the steps required to perform the simulations, defining and implementing a buffer of molecules capable of controlling the concentration of proteins in the vicinity of the NP (Vilanova, 2015).

Future research in this field of NP-corona interactions with biological systems should expand beyond the limited scope of studying the competition among just a few types of proteins. Biological media, such as blood plasma, are incredibly complex and contain over 3,700 identified proteins, leading to a highly competitive environment for the formation of the corona. This phenomenon, known as the Vroman effect, highlights the dynamic nature of protein adsorption, where proteins with higher mobility and lower affinity initially occupy the NP surface, only to be replaced by proteins with higher affinity over time.

In addition to proteins, several other molecules in whole blood can contribute to the formation of the NP corona (Lundqvist et al., 2017), including lipids, carbohydrates, nucleic acids, metabolites, complement factors, and antibodies (Soliman et al., 2024a). For example, lipids such as cholesterol and phospholipids, glycans—which are carbohydrate molecules that play a role in cell recognition and signaling—or small metabolites like glucose, hormones, and vitamins can adsorb onto NPs, altering their surface properties and influencing their interactions with cells (Singh et al., 2021).

On the other hand, circulating DNA is present in the blood of all individuals (Van Der Vaart and Pretorius, 2008), and its concentration increases in cancer patients (Jahr et al., 2001). Furthermore, depending on the state of different diseases, blood plasma also contains a variety of RNA types, including mRNAs, noncoding RNAs, and fragments of rRNAs, snoRNAs, and miRNAs (Savelyeva et al., 2017; Semenov et al., 2008), suggesting that the NP-corona composition also depends on the health condition of the host.

Moreover, complement factors that are part of the immune system and immunoglobulins or antibodies, which are glycoproteins produced by plasma cells, can mark the NP for clearance by immune cells (Singh et al., 2021). Therefore, it will be critical in the future to understand how all these components collectively form the *biomolecular* hard and soft corona, which significantly impacts the biological identity and fate of NPs in the body (Soliman et al., 2024a).

To gain a comprehensive understanding of NP behavior in biological environments, it will be crucial to investigate how these biomolecules operate under healthy or disease conditions. This approach will provide insights into real-world scenarios and how stressful conditions influence the NPs' biological identity and subsequent interactions with cells and tissues.

Moreover, the study of the dynamics and aging behavior of the biomolecular corona will be essential. As discussed here, the corona's composition and structure can change over time, exhibiting glassy behavior characterized by slow dynamics and structural rearrangements. Understanding these processes is vital for predicting the fate and stability of NPs in biological environments. The glassy state of the corona may play a significant role in determining the long-term interactions of NPs with biological systems, impacting their efficacy and safety in medical applications.

Considering the corona's glassy states is particularly important for the development of stable pharmaceuticals. Ensuring that the corona remains properly formulated when NPs are immersed in biological fluids is crucial for maintaining their intended function and avoiding unintended side effects. This knowledge can guide the design of NP-based drug delivery systems, enhancing their stability and performance in the complex biological milieu.

From a computational perspective, the scientific community will need to develop new approaches to account for the interaction of NP-corona complexes with cells, particularly focusing on their interaction with cell membranes. Recent findings indicate that the membrane interface is more complex than traditionally thought (Martelli et al., 2021). This interface extends further than previously suggested, including several water layers up to 2.5 nm (Martelli et al., 2018), and possesses an internal structure composed of both bound and unbound water (Calero and Franzese, 2019). This structure arises from the specific hydrogen-bond network of the hydration water (Bianco et al., 2012).

Current coarse-grain membrane models do not incorporate this hydrogen-bond network (Yesylevskyy et al., 2010). As a result, they fail to replicate the dynamic and thermodynamic anomalies of water (de los Santos and Franzese, 2011), which are crucial to understanding the physics of water (Leoni et al., 2021) and proteins (Bianco et al., 2017). This limitation hinders the study of interactions between the biomolecules forming the corona and the cell membrane.

Progress in this area has been made by developing a quantitatively accurate model of water under life-relevant conditions (Coronas and Franzese, 2024; Coronas et al., 2024). This model is reliable, efficient, scalable, and transferable, meeting the requirements for biological simulations. Such advancements are essential for accurately simulating the complex interactions at the NP-corona and cell membrane interface.

In summary, future studies should adopt a holistic approach that considers the vast array of molecules present in biological media and the competitive interactions that occur on NP surfaces and within the soft corona. Additionally, a deeper understanding of the dynamic and aging behaviors of the biomolecular corona, as well as the water-mediated interaction with the cell membrane, will be instrumental in advancing NP applications in medicine, leading to the development of more effective and safer nanotherapeutics.

## Data availability statement

The datasets presented in this study can be found in online repositories. The names of the repository/repositories and accession number(s) can be found in the article/supplementary material.

## Author contributions

OV: Data curation, Formal Analysis, Investigation, Methodology, Software, Validation, Visualization, Writing–original draft. AM-S: Formal Analysis, Investigation, Visualization, Writing–original draft, Writing–review and editing. MM: Conceptualization, Funding acquisition, Methodology, Supervision, Writing–review and editing. GF: Conceptualization, Formal Analysis, Funding acquisition, Methodology, Project administration, Supervision, Writing–review and editing.

## Funding

The author(s) declare that financial support was received for the research, authorship, and/or publication of this article. OV acknowledges financial support from IN2UB. AM-S and MM acknowledge the support from H2020 grant no. 952924 (SUNSHINE). GF acknowledges the support by a) MCIN/AEI/10.13039/501100011033 and "ERDF A way of making Europe" grant number PID 2021-124297NB-C31, b) the Ministry of Universities 2023-2024 Mobility Subprogram within the Talent and its Employability Promotion State Program (PEICTI 2021-2023), and c) the Visitor Program of the Max Planck Institute for The Physics of Complex Systems for supporting a visit started in November 2022. All authors acknowledge the support of NVIDIA Corporation's Applied Research Accelerator Program for granting an RTX A5000 GPU used for the calculations presented here. NVIDIA Corporation was not involved in the study design, collection, analysis, interpretation of data, the writing of this article, or the decision to submit it for publication.

## Acknowledgments

OV and GF thank Alejandro Rodríguez Ruiz and Delia López Jareño for preliminary tests of the simulation package BUBBLES.

## Conflict of interest

The authors declare that the research was conducted in the absence of any commercial or financial relationships that could be construed as a potential conflict of interest.

The author(s) declared that they were an editorial board member of Frontiers, at the time of submission. This had no impact on the peer review process and the final decision.

## Generative AI statement

The author(s) declare that no Generative AI was used in the creation of this manuscript.

## Publisher's note